\documentclass[]{article}

\usepackage{algorithm}
\usepackage{algorithmic}
\usepackage{graphicx}

\usepackage{amsfonts}
\usepackage{amsmath}
\usepackage{natbib}

\begin{document}


\title{Staff dimensioning in homecare services with uncertain demands}


\author{C. Rodriguez\thanks{Centre Micro-\'electronique de Provence Georges Charpak (SFL-ROGI, LIMOS UMR CNRS 6158), Gardanne, France}, T. Garaix$^\dag$\thanks{Centre Ing\'enierie et Sant\'e (I4S-ROGI, LIMOS UMR CNRS 6158), \'Ecole Nationale Sup\'erieure des Mines de Saint-\'Etienne, Saint-\'Etienne, France}\thanks{Corresponding author. Email: garaix@emse.fr}, X. Xie$^\dag$ and V. Augusto$^\dag$}
\date{2014}

\maketitle

\begin{abstract}
The problem addressed in this paper is how to calculate the amount of personnel required to ensure the activity of a home health care (HHC) center on a tactical horizon. Design of quantitative approaches for this question is challenging. The number of caregivers has to be determined for each profession in order to balance the coverage of patients in a region and the workforce cost over several months. Unknown demand in care and spatial dimensions, combination of skills to cover a care and individual trips visiting patients make the underlaying optimization problem very hard. Few studies are dedicated to staff dimensioning for HHC compared to patient to nurses assignment/sequencing and centers location problems.  We propose an original two-stage approach based on integer linear stochastic programming, that exploits historical medical data. The first stage calculates (near-)optimal levels of resources for possible demand scenarios, while the second stage computes the optimal number of caregiver for each profession to meet a target coverage indicator. For decision-makers, our algorithm gives the number of employees for each category required to satisfy the demand without any recourse (overtime, external resources) with fixed probability and confidence interval. The approach has been tested on various instances built from data of the French agency of hospitalization data (ATIH).

\end{abstract}

\textbf{KEYWORDS:} homecare, staff dimensioning, stochastic programming, stochastic vehicle routing problem

\section{Introduction}
\label{sec:1}

Access to health-care services is a critical challenge of the XXIst century in modern societies.  Delivering high quality care at the right moment to the population at the right cost is a priority for all health-care systems. The Institute of Medicine~\citep{reid2005building} identified time pertinence, patient-focus and efficiency as most important criteria to ensure high quality care. Since health-care budgets are limited nowadays, optimization of resources is explored to increase performance of such systems. The need to better use resources and optimize delivery, is challenged by a constant increase of costs~\citep{chahed2006operations, matta2012modelling, verjan2013economic}.

Home health-care (HHC) is a good alternative to traditional hospitalization to offer a better access to health-care services and increase patient satisfaction and quality of care~\citep{verjan2013economic}. It has been proven that health-care at home improves access in rural areas and decreases the load of hospitals. However such structures face important challenges to deliver care in rural areas while having lower costs than traditional hospitals. Complex and technical cares (such as chemotherapy) can be delivered at home using such structures. The challenge is to do it without increasing budget.

In this article, we tackle the problem of staff dimensioning in HHC structures. In traditional hospitals, all resources are gathered in the same place and traveling distances from one patient to the next one can be neglected. In HHC systems, traveling distances between patients must be taken into account to determine resource capacity and usage, especially when freelance resources (such as city based nurses, pharmacists or doctors) are not available in some regions. In addition, qualified caregivers can not be easily hired or dismissed according to the demand variation on a middle term horizon (few months) so the staff dimensioning problem is a critical tactical decision for health-care suppliers. Long-term hiring improves employees quality of life and their own skills for regular tasks. For all these reasons, economic viability of HHC structures is very sensitive to staff dimensioning and such decisions should be taken carefully by considering all parameters (demand evolution, availability of freelance resources on the territory, traveling distances...).

Uncertainties related to short term and long term decision support systems are a great challenge in many activities as supply chain management \citep{gupta2003managing, peidro2009quantitative}, transportation and logistics \citep{list2003robust} and production planning \citep{gatica2003capacity}. HHC context presents a wide variety of uncertainties compared to many industrial activities. Like in many health-care systems demand parameters are stochastic in volume, timing, pathology and severity, but also on spatial location. In addition, postponement or ignorance of patients requests are rarely feasible options for HHC. In~\cite{lanzarone2010patient}, a stochastic model of patient's care pathway in HHC allows to predict estimation of patient's requirements. But some authors~\citep{leykum2014manifestations} show that the uncertain nature of the procedures and the diseases must be taken into account in the design of the systems. They found that while process-based efforts are efficient in low-uncertainty context they are insufficient in the high-uncertainty situations. In order to prevent these drastic changes, some authors~\citep{shishebori2015robust} have studied the robust approach to protect the network against changes.  In another research~\citep{argiento2014bayesian}, these elements are introduced in the design of the system.

Staff dimensioning in traditional hospital is less critical than in HHC structures for two main reasons related to the close proximity of patients: i)~activities can easily be shared or exchanged in case of employees shortage; ii)~working time can be saved by small reductions in the time spent for each care. Classical approaches for staff dimensioning are based on calculating requirements using mean values of historical data of demand, the definition of a global budget~\citep{busby2006decision} or a given budget and a focus on improvements in service delivery~\citep{de1998planning}. Dimensioning staff is critical for small HHC structures that represent 57\% of French HHC centers where less than 10,000 hospitalization days were performed in 2011~\citep{FNEHAD2011hospitalisation}. In that case, the efficiency of the system is sensible to small changes in the number of available personnel. Detailed operational aspects (routing, mix of pathologies and mix of skills for pathologies care) are therefore integrated in our approach dedicated to HHC structures.

The main contribution of this article is an approach to size human resources of HHC structures taking into account combination of skills for each service and uncertainties related to demand evolution and location. The decision level is tactical and allows HHC managers to plan the deployment of a structure in a territory with a minimal amount of information. A two-phase approach based on a quasi-exact Monte-Carlo approach is proposed to solve this problem where the stochastic demand is estimated by a set of scenarios. These scenarios capture the complex nature of the demand over a territory and constitutes an original and efficient way to take into account medical parameters of the problem.

In order to classify our contribution, the taxonomy of~\cite{hulshof2012taxonomic} can be used. They divided the research fields in a matrix. In the vertical axis, works are divided following the planning horizon (strategic, tactical and operational decisions). In the horizontal axis the different health-care services (Ambulatory, Emergency, Surgical, Inpatient, Home and Residential). Following this classification, our contribution is situated in tactical level and the HHC service. Our approach integrates operational decisions, but restricted to the evaluation of tactical decisions. More details about the specific place of this work in the research landscape is given in the next section.

The article is organized as follows. A literature review on staff dimensioning and related problems is proposed in Section~\ref{sec:litRev}. Problem statement is given in Section~\ref{sec:2}. Section~\ref{sec:3} presents the two-phase approach to solve the staff dimensioning problem. Section~\ref{sec:exp} presents a benchmark on results and computational experiments. Finally, conclusions and perspectives are given in Section~\ref{sec:6}.

\section{Literature Review}
\label{sec:litRev}

A classification of planning problems can be found in the work of~\cite{lanzarone2012operations} where they propose to divide human resources planning in four different levels: i)~dimensioning, ii)~districting problem \citep{benzarti2013operations,blais2003solving} iii)~assignment to visits or patients \citep{boldy1980geographical} and iv)~scheduling \citep{borsani2006home} and routing \citep{begur1997integrated}. This study can be completed with the framework proposed by \cite{matta2012modelling} where authors classify management decisions from their hierarchical perspective. Dimensioning is a fundamental part of the planning system since its decision defines main constraints of the other three planning systems, the best policies of which are impacted by staff dimensioning decisions. Moreover, the capacity is almost defined by the medical or paramedical staff in HHC systems \citep{lanzarone2012operations}. This problem must be understood as a part of decision systems that are critical to the survival of the HHC structures. Literature review on staff dimensioning problem will be presented as follows. The first section is dedicated to the problem applied in health-care services. The research in other application fields is presented in the second section. This problem has been extensively studied in structures (traditional hospitals, factories, call centers etc.) where resources are not moving or where time spent to move can be easily estimated. On the contrary, literature is rare when staff routing aspects need to be considered.

\subsection{Staff planning in HHC}

Almost all works of the literature dealing with staff planning in HHC are related to some already set structures. The objective is to improve operational working. We state a lack pf studies on staff dimensioning problem for HHC. Some recent studies on assignment, scheduling and routing problems are presented here in order to better understand the research context of this paper, as these problems are strongly related to the staff dimensioning decisions. 

The districting problem divides the territory in groups of patients and health-care professionals following different criteria. The main objective is to reduce costs or improve the matching of offer and demand. In~\cite{benzarti2013operations} for example, optimization criteria are the workload balance, the maximum distance between two units and the indivisibility of assignment (one unit can only be assigned to one district). In this study, authors assume that patients have the same profile. They propose two different models and evaluate them in randomly generated scenarios. Another example can be found in~\cite{blais2003solving} where authors apply a multi-criteria Tabu-search algorithm to divide a real territory in Canada.

Another staff planning problem is nurse-to-patient assignment problem. Here, the new patients must be assigned to a specific resource while keeping some constraints. The most common constraints are, skills of resources, resource capacity and continuity of care (some patients have to be visited by the same nurse, as far as possible). In the work of~\cite{lanzarone2014robust} the authors propose a mathematical model to assign patients to nurses for health-care providers. They take into account a random demand, fixed transportation time (demand is modeled as the number of visits per week) and the objective is to minimize the maximum overtime of a resource. They present a structural policy consisting mainly on individual patient assignment (ranked by demand) while the expected cost is minimized. In~\cite{yalcindag2012operator}, the authors introduce routing considerations into the assignment. They propose an assignment-first routing-second approach on a single district. They compare two structural policies and a mixed-integer linear program for the assignment and conclude that the last approach is the best one. For routing considerations they solve a classical Traveling Salesman Problem for every resource. In the study of~\cite{carello2014cardinality} authors highlight the uncertain aspect of demand. They propose a cardinality constrained robust model to solve the nurse-to-patient assignment problem. One of the important contributions of this study is the modeling of the continuity of care. This constraint, often considered as a soft constraint, divides here patients in different subsets: enforced (during all the treatment or just at the beginning), partial and none.

The scheduling and routing problems have been extensively studied in literature. For a complete review readers are refereed to~\cite{lanzarone2012operations} and~\cite{hulshof2012taxonomic}. One recent study is~\cite{allaoua2013matheuristic} where the authors model the problem of nurse scheduling and rostering as a vehicle routing problem with time windows. In~\cite{liu2013heuristic}, the authors present two heuristics to solve the problem of sample pick-up and medicine delivery (simultaneously) with time windows. Finally, \cite{kergosien2014routing} present a problem dealing with pick-up of blood and urine samples in HHC. They solve the problem using a Tabu-search algorithm based on a variable neighborhood search.

In all the studies presented in this section, the number of caregivers is a given constant number. In the next section, works considering staff dimensioning in other contexts are presented.

\subsection{Staff dimensioning in health-care services and other industries}

A complete literature review on planning in health-care systems can be found in \cite{hulshof2012taxonomic}. The authors present a taxonomy to classify planning problems applied to health-care. Their study is very complete including up to 400 references. Here only some recent studies will be presented regarding staff dimensioning in HHC. 

Several researchers paid attention to staff dimensioning in traditional hospitals, although this problem is not as critical as in HHC. The study of~\cite{rohleder2011using} intends to control and improve patient flow in an outpatient unit. Their simulation model can be used to adjust the number of required human resources. They find that adding some key resources, like a X-ray technician, could reduce waiting times. Another study in an outpatient unit is developed by~\cite{wang2012modeling} where the authors use Markov chains modeling to improve the work flow of a tomography department in a hospital. They change the amount of staff to test different policies. 

\cite{fanti2013three} propose a three levels approach to design hospital departments. Indeed staff dimensioning takes an important part of their decision support system. The approach is based on UML (Unified Modeling Language) and Petri nets modeling, an optimization module consisting in approximating the Time Petri net model to the Time Continuous Petri net, and a simulation-decision module. They present a case study where they minimize the number of resources required to maintain the flow of patients in a hospital in Italy. 

Authors seem to have particular interests using queuing and discrete simulation models for staff dimensioning. \cite{alfonso2012modeling} propose to use Petri nets modeling linked with Discrete Event Simulation (DES) and a quantitative approach to design human resources and donor appointment strategies in blood collection systems including fixed and mobile collection sites, walk-in and scheduled donors, stochastic donor's behavior and random collection times. Another study in the blood collection system is presented in~\cite{blake2014determining} where a queuing model is used to minimize staff while maintaining waiting time requirements. It is applied in the Canadian Blood services including 51 standard models applied to 220 clinic configurations. Another case is the work of~\cite{green2006using}, where DES is used to change the amount of available staff after demand variations in an emergency department. In \cite{yom2014erlang}, a modified Erlang distribution is used to introduce the reentry of patients. A queuing model determines how many physicians/nurses are needed to maintain service levels. The constraint-programming model in \cite{ganguly2014emergency} calculates the required staff and schedules them to serve patients in an emergency department.

Dimensioning resources is an important problem in health-care, especially in HHC structures where special features such as demand distribution impact the economic pertinence and viability of the structure. A literature review can be found on approaches intended to flexible and adaptable health-care structures~\citep{carthey2010achieving}. Bed allocation in a specific clinic is considered by using queuing theory, particularly the Erlang-loss model~\citep{de2010dimensioning}. The authors develop a decision support system and, considering two-years data, they show that merging units can decrease the number of required beds. Another work based on DES and Petri Nets to plan the capacity of maternity HHC structures in Ha\"iti can be found in~\cite{norly2010dimensionnement}; authors simulate a maternity service where an additional resource is added every time a patient waits more than 50 minutes. Even if the authors identify the distance between the health-care providers and patients as the main difficulty for delivering good quality services, they do not make clear how this parameter is taken into account in their approach. Finally, another example can be found in~\cite{trilling2006aide} where the authors design a simulation model in the context of shared resources in the hospital. In order to calculate resource's workload, the simulation with infinite capacity.

An important part of staff dimensioning studies has been realized in call centers industry, where calls follow probabilistic distributions according to their type and resources have different skills -- but single assignment to calls. In this case the problem consists in minimizing the number of resources to accomplish all tasks. Markov chain models have been widely used to deal with such problems as presented in~\citep{koole2002queueing}. The problem can also be considered as a scheduling problem and can be solved using deterministic constraint programming like in~\cite{canon2005dimensioning}. Random instances are used to test the algorithm finding that optimal solution can be reached within two minutes for instances with up to 80 jobs. Shared resources are involved in the call center problem of~\cite{akcsin2003capacity}.

Resource dimensioning considering demand uncertainty has been studied in the recent literature~\citep{sahinidis2004optimization,peidro2009quantitative,guillen2005multiobjective}. A survey considering the demand uncertainty on the supply chain can be found~\citep{gupta2003managing} where the authors divide the models in two categories: i)~Manufacturing decisions ("here-and-now") and ii)~Logistics decisions ("wait-and-see"). In the pharmaceutical industry a study was made to plan the investments taking into account the uncertain result of clinical tests~\citep{gatica2003capacity}; the authors propose a multi-stage optimization problem where decisions are taken when information is revealed. The probabilities of success of a product evolves after each step of the clinical trial. Probabilities between steps of the trial are assumed related and not independent. The model is solved with Branch-\&-Bound and is capable to solve instances with three deterministic products and one stochastic. In~\cite{list2003robust}, the authors present a fleet sizing problem under uncertainties. There are two sources of uncertainty: (i)~demand and (ii)~performance of vehicles.

Almost all works cited above model systems where patients (or demands) arrive to a facility (hospital, blood collection site, outpatient units) where several resources can be assigned to each patient. Therefore, traveling times  are neglected in the definition of the capacity of the service. Another specificity of the staff dimensioning problem in health-care context comes from mix of cares required for a the pathology. For instance, if a treatment required 1 hour of nurse and 30 minutes of an oncologist, sizing decisions on both professions have to be integrated. At the operational level, planning decisions of different professions are less dependent since the demand to cover is known for each profession. That combinatorial integration increases the difficulty of the staff dimensioning problem.

In order to solve this problem we propose innovative methods with the following scientific contributions: i)~a quantitative approach to size staff of HHC with uncertainties, routing aspects and complex coverage function of demands, ii)~an original service level definition to ensure the robustness of the solution, iii)~a complex demand modeling including three sources of uncertainty (geographical, epidemiological and volume). 

In the next section the problem will be defined. In section \ref{sec:3} the two-phase approach to solve the problem will be presented. In section \ref{sec:exp} the computational experiments will be proposed and the conclusion will be discussed in section \ref{sec:6}.

\section{Problem definition}
\label{sec:2}

The key question in this study is how many human resources are necessary to attend demand. The objective of the problem is to give the minimum amount of human resources in order to cover a certain percentage of days, called  'performance level'.

The mix of demands from different pathologies, (requiring different amounts of time of different resources like nurses, doctors, specialists, etc.) is not known in advance. Demands can appear in different locations on the territory assigned to the HHC structures by the authorities. The amount of demand is also unknown and routing aspect is critical as about 30\% of working time is spent in traveling. Planning human resources without this information is complicated and the consequences of bad decisions can be important. However, some information can be used to estimate demand. In this study three main sources of information are considered to be the input of the problem: i)~geographical information around the HHC structure, ii)~historical data of demand and iii)~information of the resources required to treat the pathologies in the HHC structure.

Decision-makers and HHC planners can use the approach developed in this paper whenever they have the information previously mentioned about the territory to cover and want to know the minimum amount of personnel of each profession required to meet the target performance level. The following part of the problem description presents some modeling aspects and explain the fundamental hypothesis of the approach.

Consider a HHC structure that operates on a territory $\mathcal{T}$ divided into sectors in the set $S$. A sector $s$ corresponds to an homogeneous area where intrasectoral travel times, denoted $t_{ss}$, can be assumed constant for any pair of origin and destination belonging to sector $s$.  Between sectors $s_1$ and $s_2$, the intersectoral travel time $t_{s_1s_2}$ is constant along the optimization horizon $H$. Moreover, each sector is large enough to allow us to obtain statistical data on every pathology (ATIH~\footnote{ATIH is the French technical agency for hospitalization information, http://www.atih.sante.fr. From the ATIH database, one can get the amount of patients of every region in France for every pathology and the treatment associated with. This data is public and three years of data (from 2010 to 2012) were used in this study.} is the main source of real-life data used in this paper). This database offers epidemiology data related to every GHM~\footnote{GHM: \textit{Groupe Homog\`ene des Malade} is a classification of treatments used by French authorities to calculate reimbursements, hospital capacities, resources requirements and so on.}. We assume that patients who are suffering from the same pathology require the same care with multiple activities. The cost associated to each required employee depends on his/her profession $p$, among $P$ the set of professions. As a permanent member, an employee generates a constant cost $c_p$ for the whole optimization horizon. Each activity $a\in A$ requires $w_{a,p}$ working hours for profession $p\in P$. Note that some activities can be performed remotely and thus, do not require any move to the sector of the patient.

The demand matrix $\mathbf{d}=\left[d_{sa}\right]$ gives the daily demand for each activity at each sector. The daily demand applied to a territory is called scenario on this territory. The total number of demand $D$, for each day is defined as a random variable.
Each demand has probability $\pi_s$ to be in sector $s$ and corresponds to
activity $a$ with probability $\rho_a$. Thus, $D$ defines the demand
distribution, $\pi$ the spatial distribution and $\rho$ the epidemiological
distribution. In the following sections, these distributions are independent but
the generalization is trivial.

In practice several HHC structures use freelance resources usually located near to patients that are distant.
 But this approach can imply heavy economical charges to the structure and the use of these resources should be minimized. In our model, when HHC structure is capable of serving all patients of a certain day without using freelance resources (and without extra hours) the day is considered 'covered' and 'uncovered' otherwise the day is considered 'uncovered'. Thus, we define the level of performance $\alpha^*$ as the number of possible day (called scenario in the rest of the text) covered for a given staff.

The problem is to determine $n_p$, the number of employee for each profession $p$, such that the cost $\sum_p c_pn_p$ is minimized and the level of performance $\alpha^*$ satisfied on $H$. A scenario is covered by a solution $\mathbf{n}$ when its corresponding daily demand is completely served by resources described by $\mathbf{n}$. In order to determine if a scenario can be covered with $\left[n_p\right]$ resources, the resources assignments to activities have to be decided. Those variables are denoted $q_{sapk}(\mathbf{d})$ and represent the number of activities $a$ served by the employee $k$ of profession $p$ in the sector $s$. Thus, the total working time of $k$ can be calculated by computing the route visiting each sector where $k$ has to operate. As the daily working time is limited to $L$, for any solution $\left[n_p\right]$ the feasibility (covering) problem can be decided for each day.

\section{Two-phases approach}
\label{sec:3}

 The problem described above can be seen as a two-stage stochastic programming~\citep{birge1997introduction,dantzig1955linear}. The first stage only relates to $n_p$ decision 
variables. At the second phase, when demands are known, the assignment of patients to employees selected at the first phase is solved and their routing plan can be done. This NP-hard problem (as its deterministic version contains the well-known Vehicle Routing Problem) is difficult to address directly. Therefore, we propose a two-phases approach exploiting the definition of the performance level $\alpha^*$.
Because of the uncertainty of demand our algorithm is a quasi-exact Monte-Carlo approach where the stochastic demand is estimated by a set of scenarios.

\begin{figure}[h]
\begin{center}
\includegraphics[scale=0.9]{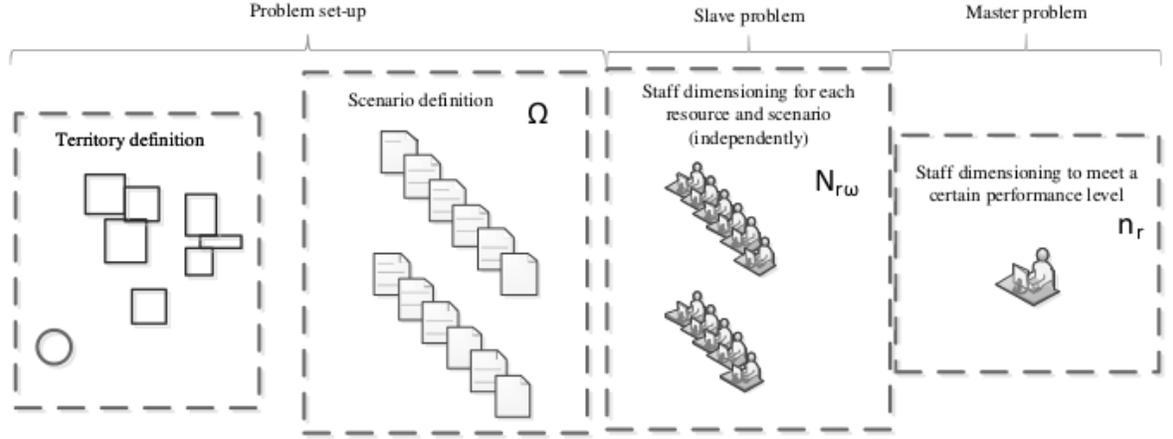}
\caption{The two-phases approach.}
\label{fig:approach}
\end{center}
\end{figure}

In order to evaluate the performance of a solution $\mathbf{n}=\left[n_p\right]$, an optimization problem can be solved maximizing the number of demands served. Another approach minimizes overtimes needed to cover all the demands of each scenario for each profession. The approach proposed in this paper (see Figure~\ref{fig:approach}) allows to decompose the problem for each profession, since daily activities done by different professions are assumed independent. But it implies to solve the routing problem for each solution.
Our approach tries to find the minimal number of resources $N_{p\omega}$ of profession $p$ required to serve all
the demands of each scenario $\omega$. This daily problem is called the slave problem $SP(p,\omega)$ and does not depend on solutions $\mathbf{n}$.

As $N_{p\omega}$ values are computed, a scenario is covered by a solution $\mathbf{n}$ if and only
if $n_p\geq N_{p\omega}$ for each profession $p$. Now, the staff dimensioning
can be modeled by the so-called master problem defined with those constraints.

Our approach to solve the staff dimensioning problem of a territory has three steps:
\begin{enumerate}
 \item\label{e:3_1} generate a set $\Omega$ of daily scenarios of demands;
 \item\label{e:3_2} compute $N_{p\omega}$ for each scenario $\omega$ and each profession $p$ by solving
 the slave problems $SP(p,\omega)$.
 \item\label{e:3_3} solve the master problem and get the optimum $\mathbf{n}=\left[n_p\right]$.
\end{enumerate}

The three steps are detailed in the three following sections. Then some modeling improvements are described in Section~\ref{ssec:3_3}.

\subsection{Generation of scenarios}
\label{ssec:3_0}

 Each scenario corresponds to a possible daily demand, generated from a territory. All data that define a territory, are listed in Section~\ref{sec:2}. They contain all characteristics of sectors, activities and professions. For each territory, some demand patterns are generated on three fields: the total number of demands ($=\sum_{(s,a) \in S\times A} d_{sa}$), probabilities of any demand on localization ($\pi_s, \forall s\in S$) and on its nature ($\pi_a, \forall a\in A$).

\subsection{Slave problem}
\label{ssec:3_2}

Before solving the master problem, minimal resource requirements $N_{p\omega}$ have to be computed for each scenario $\omega$ and each profession $p$. In the sake of clarity, $p$ and $\omega$ are omitted in this section.

The schedules of each resource have to be optimized in order to decide the minimal number of resources necessary to serve all. Each required resource has to visit some patients and also to serve some remote activities.

The slave problem is modeled as an integer linear program described below. This model is based on route variables, and so comes with a huge number of variables. In Section~\ref{ssec:3_3}, problem properties are exhibited. They allow to reduce the number of variables to a tractable size for the generated scenarios.

An integer linear program ensures the demand covering. In order to limit the size of the problem, routes represent sequences of sectors and do not consider cares and demands. Then the assignment of the demands to vehicles at each sector has to be explicitly decided in this model. It differs from the unique paper we found on split deliveries with discrete quantities \citep{Sal11}. In their model, the routes also define the quantities delivered at each position. Because of the large number of such routes, a branch-and-price scheme is used and takes advantage of other constraints of their problem like the time window constraints. A covering model with routes as sequences of nodes without undefined quantities has also been used by~\cite{Arc12} for an inventory routing problem, where stock level, delivering and routing decisions are integrated. They select, in that way, routes among a set of routes derived from solutions obtained by a Tabu Search algorithm. Their problem is easier since quantities delivered are real numbers.

A dummy sector $s=0$ is associated to the HHC structure building. Without loss of generality, all the demands of remotely served activities for the considered profession are transferred to sector $0$. Set $S$ and the demands $\mathbf{d}=\left[d_{as}\right]$ are accordingly updated and the intrasector duration $t_{00}$ is set to $0$.

Let consider $K$ the set of resources and $R$ the set of feasible routes that can be assigned to resources $k\in K$. Schedules associated to resources can be modeled through binary $x_{kr}$ and integer $q_{ask}$ decision variables, that represent assignments of route $r$ and proportion of activity $a$ in sector $s$ to resource $k$, respectively. The length of each route $r$ is known and denoted $T_r$ that corresponds to the sum of the intersector travel durations. Note that the node representing the HHC structure building is included in all routes. It also gives single node routes modeling employee who exclusively works remotely. Additional binary parameters $a_{rs}$ indicate whether route $r$ is passing through sector $s$. Thus, the slave problem is given by equations~(\ref{e:sp0})-(\ref{e:sp6}).

\begin{eqnarray}
\min N = \sum_{(r,k)\in R\times K} x_{rk} \label{e:sp0}\\
 \textrm{subject to} \nonumber\\
 \sum_{r\in R} x_{rk} \leq 1, & \forall k\in K \label{e:sp1}\\
 \sum_{k\in K} q_{ask} \geq d_{sa},& \forall (s,a)\in S\times A \label{e:sp2}\\
 \sum_{(s,a)\in S\times A} (w_a+t_{ss})q_{ask} \leq \sum_{r\in R} (L-T_r) x_{rk},& \forall k\in K  \label{e:sp3}\\
 \sum_{a\in A} (w_a+t_{ss})q_{ask} \leq \sum_{r\in R} (L-T_r) a_{rs}x_{rk},& \forall (k,s)\in K\times (S\setminus \{0\})  \label{e:sp4}\\
 x_{rk} \in \{0,1\},& \forall (r,k) \in R\times K  \label{e:sp5} \\
 q_{ask} \in \mathbb{N},& \forall (a,s,k)\in A\times S\times K  \label{e:sp6} 
\end{eqnarray}

The objective function~(\ref{e:sp0}) gives the number of resources involved in the optimal solution by minimization of the number of routes assigned to resources. We recall that resources can only perform one route, as constraints~(\ref{e:sp1}) ensure. All demands have to be covered as in constraints~(\ref{e:sp2}). Individual capacity constraints are satisfied with~(\ref{e:sp3}) inequalities, where the left member expresses the workload -- in terms of service durations -- assigned to the resource $k$ and the right side gives the available working time minus the total travel duration. Constraints~(\ref{e:sp4}) enforce resources to pass through all sectors of the territory where they have at least one demand to serve -- if $a_{rs}$ is null then $q_{ask}$ is null. These constraints are meaningless for the dummy sector $s=0$ that is 'visited' by all resources.

\subsection{Master problem}
\label{ssec:3_1}

Given a set $\Omega$ of scenarios $\omega$, the master problem aims to find out the minimal number of resources for each profession required to cover a ratio $\alpha^*$ of scenario among $\Omega$.

The master problem is modeled as the linear integer program (\ref{e:mp0})-(\ref{e:mp4}). In addition to $n_p$ decision variables, intermediate $y_\omega$ binary variables are introduced and express if scenario $\omega$ is covered ($y_\omega=1$) or not ($y_\omega=0$) in the optimal solution.

\begin{eqnarray}
 \min \sum_{p\in P} c_pn_p \label{e:mp0}\\
 \textrm{subjet to} \nonumber\\
 N_{p\omega}y_{\omega} \leq n_p,& \forall \omega\in \Omega, p\in P \label{e:mp1}\\
 \sum_{\omega\in \Omega} y_\omega \geq \alpha |\Omega|& \label{e:mp2}\\
 y_{\omega} \in \{0,1\},& \forall \omega\in \Omega  \label{e:mp3}\\
 n_{p} \in \mathbb{N},& \forall p\in P  \label{e:mp4}
\end{eqnarray}

The objective function (\ref{e:mp0}) minimizes the cost of staff involved in the optimal solution. Constraints (\ref{e:mp1}) enforce to select more resources than each selected scenario requires, for each profession. The ratio of covered scenarios is ensured by constraints (\ref{e:mp2}) with parameter $\alpha$ (instead of $\alpha^*$), the definition of which is given in the next paragraph. The number of resources selected can take integer values with constraints (\ref{e:mp4}).

As $\Omega$ has a limited size in practice, a solution satisfying the ratio $\alpha$ on a sample $\Omega$ has a small probability to satisfy the performance level $\alpha^*$ over all possible scenarios. Let define the average observed coverage ratio $\bar{\alpha}=\sum_{\omega\in \Omega} y^*_\omega \div |\Omega|$, where $y^*_\omega$ are optimal values of the model~(\ref{e:mp0})-(\ref{e:mp4}). If we consider the random selection of a daily scenario as a Bernouilli trial, then the sum of results of such trials (\textit{i.e.}, $\sum_{\omega\in \Omega} y^*_\omega$) follows a binomial distribution that we can approximate by a normal distribution -- we assume large enough $\Omega$ and $\bar{\alpha}$ far enough from 0 and 1. The lower bound on the confidence interval at the 95\% confidence level is computed as $\hat{\alpha}-1.66 \sqrt{\hat{\alpha}(1-\hat{\alpha})}\div \sqrt{|\Omega|}$, where the observed variance is under the first square root and the number of samples under the second one. The parameter $1.66$ comes from the Student-t distribution table for the above confidence interval. 

The value $\alpha$ is experimentally set such that the obtained value $\bar{\alpha}$ gives a lower bound equal to (or slightly greater than) $\alpha^*$.

For instance, with $|\Omega|=100$ and target $\alpha^*=0.80$, we experiment that $\alpha=0.86$ gives solutions that cover $86\%$ ($\bar{\alpha}=0.86$) of scenarios of $\Omega$ and therefore gives a lower bound on the confidence interval close to $0.80\approx 0.86-1.66\sqrt{0.86(1-0.86)}\div \sqrt{100}$.

The computational complexity of the master problem is let as an open problem. If the number of professions and the maximal number of resources are assumed constant, then a full enumeration can be done in polynomial time (in the number of scenarios). In our case, $|P|$ and $n_p$ have small enough upper bounds to guarantee an efficient solution by a linear programming branch-and-cut algorithm.

\subsection{Modeling and algorithm improvements}
\label{ssec:3_3}

As the number of variables involved in the model~(\ref{e:sp0})-(\ref{e:sp6}) can be huge, this model has to be refined by means of filter on variables, tight bounds and cutting rules embedded in an algorithm involving all sub-problems.

\subsubsection{Size reduction of the slave problem}
\label{sssec:3_3_1}

The number of variables involved in the model~(\ref{e:sp0})-(\ref{e:sp6}) can be decreased by exploiting optimal solution properties and filtering sequences of nodes to consider or limiting the number of possible resources.

\begin{enumerate}

  \item As the set of assignable routes is the same for every resources, variables $x$ can be replaced by $|R|$  integer variables indicating the number of times that a route is served (\textit{i.e.}, index $k\in K$ can be ignored). But such variables do not to directly derive feasible assignment of resources to activities because of integrity of activities. That makes such modeling less efficient than the proposed one.

  However, the number of variables can be reduced because of optimal solutions properties. Let consider that problem~(\ref{e:sp0})-(\ref{e:sp6}) is feasible, then an optimal solution exists which routes satisfy the following properties.
  \begin{enumerate}
    \item Routes correspond to lowest cost -- as the sum of intersector travel times -- Hamiltonian cycle on the subset of sectors in each route. Thus the number of routes involved in the model~(\ref{e:sp0})-(\ref{e:sp6}) is less than $2^{|S|}$.
    \item An upper bound on the minimal duration of each route can be compared to $L$, the maximal route duration. This upper bound adds the inter and intrasectors travel times and the lowest demand duration that can be served at each visited sector.
  \end{enumerate}
  The first property is easily implemented in the dynamic programming algorithm that computes routes. A preprocessing procedure filters routes according to the second property. Note that these properties are scenario independent. The upper bound has been lightly weakened for that; all activities are considered at each sector even if not present for all scenarios.
  
  \item The number of resources $|K|$ has to be large enough to ensure feasibility but as small as possible for compactness. An upper bound, $UB_{pw}\geq |K|$, is computed by a simple constructive heuristic for each pair of profession and scenario $(p,\omega)$. For each profession, the algorithm starts with an empty route of a resource $k$ ($K(p)=\{k\}$). For each sector and each activity (selected in an arbitrary order), demands are pushed at the end of the current route. When no more demand can be assigned to the resource because of capacity constraints, another empty route is considered for an additional resource which is added to $K(p)$. 
 
\end{enumerate}

 \subsubsection{A cutting rule in the solution algorithm over all scenarios}
\label{sssec:3_3_2}
 
 Some slave problem runs can be avoided. Let us suppose that 100 scenarios have been generated and that $\alpha=0.8$. Then, at most 20 scenarios are not covered by any optimal solution of the master problem. Let $\omega'$ be ranked at the 21th position when scenarios are ordered by decreasing resource ($p$) requirements ($N_{p\omega}$). As covered scenarios $\omega$ by a solution satisfy $N_{p\omega}\leq N_{p\omega'}$, $N_{p\omega}$ can be replaced by $N_{p\omega'}$. Thus, when more than 21 slave problems of resource $k$ have been solved, $\omega'$ can be determined, and $N_{p\omega'}$ used as a lower bound, denoted $LB_{p}$ in next slave problems. All steps of the sequential solution of slave problems are detailed in Algorithm~\ref{algo:solve_sous_probleme}. This algorithm can easily be parallelized according to pairs of professions and scenarios.

The lower bound maintained in Algorithm~\ref{algo:solve_sous_probleme}, is added to model~(\ref{e:sp0})-(\ref{e:sp6}) with constraint~(\ref{e:sp8}). Actually, solutions involving less than $LB_p$ resources can be replaced by $LB_p$ in the master problem. As constraints~(\ref{e:sp2}) are inequalities, solutions with exceeding resources are accepted.

\begin{equation} \label{e:sp8}
	N \geq LB_p
\end{equation}
  
\begin{algorithm}
\begin{algorithmic}
\REQUIRE profession $p$, territory $\mathcal{T}$
\STATE $LB_p:=0$
\STATE $\{UB_{p\omega}\}_{\omega\in\Omega}$: the set of upper bounds on $NP_{p\omega}$
\STATE $\mathcal{L}$: the list of $\omega\in\Omega$ sorted by decreasing $UB_{p\omega}$ value
\WHILE {$\mathcal{L}\neq\emptyset$}
	\STATE $\omega := pop(\mathcal{L})$
	\IF { $LB_p \geq UB_{p\omega}$ }
	  \STATE $N_{p\omega} = LB_p $
	\ELSE 
		\STATE $N_{p\omega} :=$ best solution of (\ref{e:sp0})-(\ref{e:sp6}) with $\vert K \vert = UB_{p\omega} -1$
		\IF {(\ref{e:sp0})-(\ref{e:sp6}) is infeasible}
			\STATE $N_{p\omega} := UB_{p\omega}$
		\ENDIF
	\ENDIF
	\IF { $|\mathcal{L}| < \alpha|\Omega|$ }
		\STATE $LB_p := \max\{N_{p\omega}:|\{\omega^{\prime}: N_{p\omega^{\prime}}>N_{p\omega}\}| = (1-\alpha)|\Omega|\}$
	\ENDIF
\ENDWHILE
\end{algorithmic}
\caption{How to compute $N_{pw}$ for profession $p$ over $\Omega$}
\label{algo:solve_sous_probleme}
\end{algorithm}

\subsubsection{A lower bound for the master problem}
\label{sssec:3_3_3}

The replacement of $N_{pw}$ by lower values gives a lower bound to the master problem. When some slave problems are not solved to optimality, a lower bound can still be derived for the master problem. This lower bound is obtained by considering a model similar to~(\ref{e:mp0})-(\ref{e:mp4}), where $N_{pw}$ is replaced by the lower bound on its value given by the branch-and-cut procedure which solves each slave problem.

\section{Computational experiments}
\label{sec:exp}

In this section, we show how our approach can be used to determine the level of staff required to cover a territory in Section~\ref{ss:qualitative}. In Section~\ref{ss:compare}, we show the benefit of our approach against less sophisticated computational methods. The benchmark used are described in Section~\ref{ssec:bench}. The settings, the performance and the limits of our algorithm are discussed in Section~\ref{ss:performance}.

Master and slave problems are solved by the parallel branch-and-cut algorithm of IBM Ilog Cplex~12.5 in a C++ code. All runs have been performed on a 8 processor Linux machine rated at 2.4 MHz. A running time limit of 300 seconds has been imposed to each call to the slave problem solution.

\subsection{Benchmark}
\label{ssec:bench}

An instance is defined by a territory, a demand pattern and a set of scenarios (daily demands). 

This information is built from historical data on pathologies and cares. Only four different types of care are modeled; they correspond to the three main cares over 24 plus a dummy care that represents an average of less frequent ones. The three most representative cares are: palliative cares ($\pi_a\approx 0.26$), complex bandage and ostomy ($\pi_a\approx 0.23$), and heavy nursing cares ($\pi_a\approx 0.10$). According to ATIH data (2012) provided by home care services in France, more than one half of day of hospitalization at home is covered by these three type of cares. Data recorded by the ATIH in 2012 involve 317 public and private HHC structures reporting 4 207 177 days of hospitalization and an average length of stay of 16.21 days. All French HHC structures have to send to ATIH their activities detailed per type of care (GHM) and type of employee.

The time spent by resources to care patients is reported in Table~\ref{t:duration} for nurses, nurses' aid and physicians. Durations in the last line result from a weighted sum of durations of pathologies ; weights are frequencies. All durations are estimated from ATIH data. Other workers are discarded, since they are parsimoniously used and certain can operate as subcontractors of the HHC structure. Note that the impact of the increase of the number of cares, can be reduced by merging cares with the same service durations for each type of caregiver. This trick has not been exploited in the experiments described in this section. The increase in the number of types of caregiver involved in the model linearly increases the difficulty of the problem, because of the independence of daily covering problems related to each type of caregiver.

\begin{table}
 \begin{tabular}{|l||r|r|r|}
  \hline
cares	&	nurse	&	nurse's aid	&	physician	\\	\hline
palliative	&	60	&	35	&	10	\\	
complex bandage	&	40	&	15	&	10	\\	
heavy nursing	&	45	&	50	&	10	\\	
others	&	40	&	25	&	5	\\	\hline
monthly cost (euros) & 1200 & 800 & 2500 \\\hline
\end{tabular}
 \caption{Service durations in minutes}
\label{t:duration}
\end{table}

Every demand matrix $\mathbf{d}=\left[ d_{sa}\right]$ is generated following a demand pattern on a territory, and gives one instance.

The set of 96 instances is built according to parameters listed in Tables~\ref{t:attributes}, \ref{t:territory}, \ref{t:patterns} and \ref{t:instances}. Each territory is defined by sparsity and division attributes. Sparsity parameters are detailed in Table~\ref{t:territory} where intrasector distances are uniformly generated from intervals, and intersectors distances are computed as euclidean distances from uniformly generated points in a square. In the case of semi-urban territories, two squares have the same center and the probability to be in the smaller one is 0.5. In order to be consistent with the sector-based design, intrasectors distances are generated smaller than intersectors ones.

 For each combination of sparsity and division attributes, in Table~\ref{t:attributes}, two territories are generated, to a total of 12 territories. Then, instances are generated for each territory following patterns of Tables~\ref{t:patterns} and~\ref{t:instances}. In series S1, the total amount of demand is fixed and all sectors are similar regarding the volume and the nature of the demand. In series S2, the total demand can vary. Series S3, correspond to HHC structures which allocate clusters of close demands to specific days. Finally, when the demand can not be smoothly spread on the planning horizon, some typical scenarios with different levels of demands and spatial distributions are generated, like in series 4.

\begin{table}
 \begin{tabular}{|l||c|c|c|c|}
  \hline
  \multicolumn{1}{|c||}{attribute} & \multicolumn{4}{|c|}{values}\\\hline
  Sparsity & \multicolumn{4}{c|}{rural; urban; semi-urban} \\\hline
  Division & \multicolumn{4}{c|}{10 sectors; 15 sectors} \\\hline 
  Demand pattern & \multicolumn{4}{c|}{stable; volume variation; geographical variation; typical days} \\\hline
 \end{tabular}
 \caption{Instances attributes}
\label{t:attributes}
\end{table}

\begin{table}
 \begin{tabular}{|l||r|r|}
  \hline
sparsity & inter-sector distances  & sectors location \\ \hline
rural & $[5, 15]$ & $90\times 90$ square \\
urban & $[5, 10]$ & $60\times 60$ square \\
semi-urban & $[5, 10]$ & $90\times 90$ and $60\times 60$ squares \\\hline
\end{tabular}
 \caption{Spatial design of territories}
\label{t:territory}
\end{table}

\begin{table}
 \begin{tabular}{|l||r|r|r|r|}
  \hline
  \multicolumn{1}{|c||}{demand pattern} & \multicolumn{1}{|c|}{description}\\\hline
  stable & the daily total demand is fixed \\\hline
  volume variation & the daily total demand is uniformly ranged in an interval \\\hline
  geographical variation & the daily total demand is uniformly ranged in an interval \\\hline
  typical days & 4 typical patterns of total demand and demand distribution are defined \\\hline
 \end{tabular}
 \caption{Demand patterns}
\label{t:patterns}
\end{table}

\begin{table}
 \begin{tabular}{|l|r|r|r|r|}
  \hline
  name & \multicolumn{1}{|c|}{demand pattern} & total demand & distribution ($\pi_s$) & \# instances\\\hline
  S1.1 & stable & 40 & common function & 12 \\
  S1.2 & stable & 50 & common function & 12 \\\hline  
  S2.1 & volume variation & unif(45,60) & common function & 12 \\ 
  S2.2 & volume variation & unif(30,60) & common function & 12 \\\hline  
  S3 & geographical variation & unif(40,50) & 1 out of 5 subregion represents & 24 \\
  &  &  & 80\% of one daily demand &  \\\hline  
  S4 & typical days & \{45,55,65,75\} & 4 patterns & 24 \\\hline  
 \end{tabular}
  \caption{Instances}
\label{t:instances}
\end{table}

For each instance, 100 hundred scenarios ($|\Omega|=100$) are generated. According to the master problem definition, $\alpha$ is set to 0.86 to enforce 0.80 (see Section 4.3. for calculation details) as covering ratio with a confidence greater than 95\%. Preliminary tests -- performed on S4 instances -- show that setting $\alpha$ to 0.86 gives solutions with an average coverage value close to 0.86. We accept that solutions founds are normally distributed centered at 0.86.

The number of routes ($|R|$) considered in the slave problem depends on the territory. Values in Table~\ref{t:routes}, show the efficiency of the filtering procedures applied to variables of model~(\ref{e:sp0})-(\ref{e:sp8}), as 921.0 has to be compared to $2^{10}$ (the maximal number of routes with 10 sectors) and 12295.0 to $2^{15}$.

\begin{table}
 \begin{tabular}{|l|l|l|r|}
  \hline
name & sparsity 	&	divisions	 &	$|R|$	\\	\hline
RU10 & rural	&	10	&	509.0	\\	
SU10 & semi-urban	&	10	&	633.0	\\	
UR10 & urban	&	10	&	921.0	\\ \hline	
RU15 & rural	&	15	&	3771.5	\\	
SU15 & semi-urban	&	15	&	6235.0	\\	
UR15 & urban	&	15	&	12295.0 \\	\hline  
 \end{tabular}
  \caption{Number of possible routes}
\label{t:routes}
\end{table}

\subsection{Qualitative analysis}
	\label{ss:qualitative}

For each covered scenario we consider as a day off, any resource assigned to the solution but not required by the scenario. For each covered scenario and profession, the total travel and idle times have been computed. Idle time only involves resources used in the scenario. Average values of these indicators are reported in Table~\ref{t:workload} for all series. Travel and idle times give realistic values compared to the activity of the HHC service of Sallanches. Column '\% day off' shows that the stability of the volume of demand strongly impacts the stability of the number of resources required each day.

\begin{table}[h]
 \begin{tabular}{|l||r|r|r|r|r|}
  \hline
series	&	\% day off	&	\% travel	&	\% idle	& staff & cost \\	\hline
S1.1	&	5.8	&	35.7	&	3.1	&	26.1	&	29591.6	\\	
S1.2	&	5.9	&	35.2	&	2.9	&	31.8	&	35958.3	\\	
S2.1	&	10.9	&	34.5	&	3.0	&	35.2	&	39699.9	\\	
S2.2	&	19.5	&	33.6	&	3.0	&	33.3	&	37666.4	\\	
S3	&	11.2	&	36.5	&	2.3	&	30.5	&	34683.2	\\	
S4	&	20.7	&	33.7	&	3.0	&	43.6	&	49295.7	\\	\hline
average	&	13.3	&	34.9	&	2.8	&	34.3	&	38859.2	\\	\hline
\end{tabular}
  \caption{Percentage of workload activities}
\label{t:workload}
\end{table}

\subsection{Comparison to straightforward solutions}
\label{ss:compare}

With the results reported in Tables~\ref{t:resources} and ~\ref{t:cover}, we aim to show the interest of our approach compared to trivial computations. Let us denote $inf_p$ (resp. $sup_p$) the minimum (resp. maximum) quantity of resources with profession $p$ required to serve at least one scenario of $\Omega$. From the second and third columns of Table~\ref{t:resources}, we know the average gaps between trivial bounds and the best solution ($n_p^*$) found with $\alpha=0.8$. Detailed results by instances also show that such gaps vary a lot depending on instances. On the other hand, the lower bound ($LB_p$) computed for each profession in Algorithm~\ref{algo:solve_sous_probleme} is a better approximation of the optimal solution as we can see in the last column of Table~\ref{t:resources}.

\begin{table}
 \begin{tabular}{|l||r|r|r|}
  \hline
 series	&	$\sum_p n_p^*-inf_p$	&	$\sum_p sup_p-n_p^*$	& $\sum_p n_p^*-LB_p$	\\	\hline
S1.1	&	5.3	&	1.9	&	0.3	\\	
S1.2	&	6.6	&	2.7	&	0.5	\\	
S2.1	&	10.2	&	3.3	&	0.9	\\	
S2.2	&	15.3	&	3.9	&	0.5	\\	
S3	&	9.0	&	3.5	&	0.4	\\	
S4	&	19.2	&	4.1	&	0.8	\\	\hline
average	&	11.7	&	3.4	&	0.6	\\	
maximum	&	23	&	8	&	2	\\	\hline
 \end{tabular}
  \caption{Variance of the number of resources required}
\label{t:resources}
\end{table}

Let $n^*$ denote the best solution found of the master problem.
In Table~\ref{t:cover}, the three last columns correspond respectively: the average percentage of scenarios covered by $n^*$; the average percentage of scenarios covered by $n$ but not by $n^1$ defined by formula~(\ref{e:n1}); the average percentage of scenarios covered by the solution $n^2$ defined by formula~(\ref{e:n2}) and not by $n^*$.
Solution $n^1$ is directly defined by the values of $LB_p$, the lower bound obtained by Algorithm~\ref{algo:solve_sous_probleme}. In Solution $n^2$, each scenario covered by $LB_p$ for at least one profession is covered.

\begin{equation}\label{e:n1}
	n^1 = \{n_p: p\in P, n_p = LB_p\} 
\end{equation}

\begin{equation}\label{e:n2}
	n^2 = \{n_p: p\in P, n_p = \max_{w\in \Omega} \{N_{pw}: \exists p'\in P \text{ such that } N_{p'w}\geq LB_p'\}\} 
\end{equation}

Results reported in Table~\ref{t:cover} show that the solution of the master problem is not trivial. In order to satisfy the coverage constraint, the set of scenario to cover can not be easily determined.

\begin{table}
 \begin{tabular}{|l||r|r|r|}
  \hline
series	&	$n^*$ & $n^*\setminus n^1$ 	&	$n^2\setminus n^*$ 	\\	\hline
S1.1	&	92.4	&	2.3	&	7.6	\\	
S1.2	&	89.8	&	4.3	&	10.2	\\	
S2.1	&	89.3	&	5.8	&	9.9	\\	
S2.2	&	89.3	&	3.3	&	9.0	\\	
S3	&	89.8	&	2.0	&	9.6	\\	
S4	&	87.8	&	3.4	&	11.1	\\	\hline
average	&	89.5	&	3.3	&	9.8	\\	\hline
 \end{tabular}
  \caption{Scenario coverage (\%) of trivial and optimized solutions}
\label{t:cover}
\end{table}

We also compute a Pareto pseudo-optimal set on both criteria the objective cost of resources and the percentage of coverage. As we know lower an upper bounds on the number of resources required to cover each scenario with each resource, it is easy to get all not-dominated solutions the corresponding bi-criteria master problem from solutions of all the slave problems. In order to get more accurate resources levels for each scenario we ran the program with $alpha$ set to $1$. But, we did not solve all slave problems to optimality and we therefore an approximation of the Pareto set. 

For each instance, we find a quite small (12 solutions in average) set of not-dominated solutions. Thus, obtained solutions for a fixed parameter $\alpha$ may exceed the $\alpha$ level of coverage. One can verify this statement by checking Column ``$n^*$'' in Table~\ref{t:cover}.

\subsection{Algorithm performance}
\label{ss:performance}

General performances of our approach are given in Table~\ref{t:perfcomplete}. Note that average values in the last row of tables are weighted by the number of instances per series. As Column '\% call slave' shows, about only 20.3\% require a call to the branch-and-cut solver. It highlights the interest to sort scenarios according to a quickly computed upper bound in Algorithm~\ref{algo:solve_sous_probleme}, in order to skip some slave problem solutions. But, according to Column '\% opt', optimality is not reached for more than one half of these calls within the 300 seconds limit. It means that for about 11\% (55.4\% over 20.3\%) of the scenarios, the upper bounds used on $N_{pw}$ can provide not optimal values. As 33 slave problems are not solved to optimality in average -- each scenario is evaluated for each resource --, those runs of 300 seconds represent almost all the total computing time given in Column 'cpu (h.)'.

Average and maximal gaps (restricted to the 11\% runs that do not reach optimality and computed on the staff size) indicate that the minimal number of resources required for some scenarios is difficult to evaluate, as for S1.2 and S4 the maximal gap reaches 4 resources. However, the relative average gap -- also restricted to runs that do not reach optimality --  is less than 1.2\% on all series. As described in Section~\ref{sssec:3_3_3}, the lower bound computed at each run stopped early, allows to compute a lower bound on the master problem. The average relative gap is given for every series in Column '\% gap' with an overall average value of 3.1\%. It seems that this gap can be reduced by controlling the gap obtained on each run of the slave problem.

Series S2 seem to provide more difficult slave problems than series S1. In series S2, the demand vary and can reach high values (60 demands), that generally makes the daily scenarios harder to solve.
When 80\% of the demand is allocated to 20\% of the sectors, as in series S3, more than 70\% of the call to the slave problem reach optimal solutions. Series S4, where the daily scenarios are the most heterogeneous, our approach fails to find optimum solutions to many slave problems.

\begin{table}
 \begin{tabular}{|l||r|r||r|r|r|r|}
  \hline
  & \multicolumn{2}{|c||}{Algorithm~\ref{algo:solve_sous_probleme}} & \multicolumn{4}{c|}{Slave problem} \\
series	&	\% gap	&	cpu (h.)	&	\% call slave	&	\% opt.	&	avg gap	&	max gap	\\	\hline
S1.1	&	2.3	&	2.5	&	23.5	&	66.3	&	1.1	&	2	\\	
S1.2	&	2.7	&	3.8	&	23.8	&	44.4	&	1.2	&	4	\\	
S2.1	&	3.4	&	3.7	&	19.4	&	30.0	&	1.3	&	3	\\	
S2.2	&	3.1	&	3.2	&	17.8	&	36.0	&	1.2	&	3	\\	
S3	&	2.5	&	1.9	&	21.3	&	72.4	&	1.1	&	3	\\	
S4	&	4.1	&	3.9	&	17.7	&	17.7	&	1.6	&	4	\\	\hline
average	&	3.1	&	3.1	&	20.3	&	44.6	&	1.2	&	3.1	\\	\hline
 \end{tabular}
  \caption{Computational performances}
\label{t:perfcomplete}
\end{table}

\section{Conclusion}
\label{sec:6}

In this paper we propose an original approach to estimate the staff in a HHC service. Our approach deals with a coverage constraints on demand forecast. It is useful to remind that staff dimensioning is a critical challenge for HHC structures since a small change in the number of resources will have an important impact on the economic performance (and survival) of them. Our approach allows to determine the necessary amount of caregivers using some given some data: historical demand (geographical and pathological), territory model (to estimate traveling time between sectors and inside them) and amount of time of each profession required to treat each pathology. This information can be obtained in public health-care data bases such as the ATIH.

Time spent in traveling (until one third of the working time) is taken into account. Combinations of required skills involved to cover the demand are enforced. The robustness of the approach is obtained by a scenario-based model.
This potentially very large problem is solved through an original decomposition framework. Contrary to the set of specific constraints to HHC structure, our solution framework could be applied to other combinatorial optimization problems and can be an alternative to classical stochastic programming or chance-constrained methods.
The analysis of our numerical experiments shows that routing evaluation can help to get more precise working time, especially in geographical areas assuming rural or semi-urban patterns.

Comparison to straightforward solutions and the Pareto optimal sets computed, indicate that our approach can help decision-maker before opening a HHC service or before hiring (dismissing) an employee.

\bibliographystyle{tPRS}

\end{document}